\documentclass[aip, jmp, bmf, sd, rsi, amsmath,amssymb,preprint, reprint,author-year,author-numerical,Conference Proceedings]{revtex4-1}
\usepackage{epsfig}
\usepackage[normalem]{ulem}
\usepackage[utf8]{inputenc}
\usepackage{graphics}
\usepackage{lipsum}
\usepackage{amsmath}
\usepackage{sidecap}
\usepackage{mathtools}
\usepackage{graphicx} 
\usepackage{dcolumn} 
\usepackage{bm}
\usepackage{xcolor,colortbl}
\usepackage{color, colortbl}
\definecolor{darkgray}{rgb}{0.66, 0.66, 0.66}
\definecolor{yellow-green}{rgb}{0.6, 0.8, 0.2}
\definecolor{deeppink}{rgb}{1.0, 0.08, 0.58}
\definecolor{darkviolet}{rgb}{0.58, 0.0, 0.83}

\definecolor{darkcyan}{rgb}{0.0, 0.55, 0.55}
\usepackage{sidecap}  
\begin{document}

\title{Effect of interlayer exchange coupling in spin-torque nano oscillator}
\author{R.~Arun$^{1}$}
%\email{arunbdu@gmail.com}
\author{R.~Gopal$^{2}$}
%\email{gopalphysics@gmail.com}
\author{V.~K.~Chandrasekar$^{2}$}
\email{chandru25nld@gmail.com}
\author{M.~Lakshmanan$^1$}
%\email{lakshman.cnld@gmail.com}
	
\affiliation
{
$^{1}$Department of Nonlinear Dynamics, School of Physics, Bharathidasan University, Tiruchirapalli-620024, India\\
$^{2}$Department of Physics, Centre for Nonlinear Science \& Engineering, School of Electrical \& Electronics Engineering, SASTRA Deemed University, Thanjavur- 613 401, India. \\
}

\date{\today}
\begin{abstract}
The dynamics of the magnetization of the free layer in a spin-torque nano oscillator (STNO) influenced by a noncollinear alignment between the magnetizations of the free and pinned layers due to an interlayer exchange coupling has been investigated theoretically.  The orientations of the magnetization of the free layer with that of the pinned layer have been computed through the macrospin model and they are found to match well with experimental results. The bilinear and biquadratic coupling strengths make the current to  switch the magnetization between two states or oscillate steadily. The expressions for the critical currents between which oscillations are possible and the critical bilinear coupling strength below which oscillations are not possible are derived. The frequency of the oscillations is shown to be tuned and increased to or above 300 GHz by the current which is the largest to date among nanopillar-shaped STNOs.
\end{abstract}
%\pacs{ 05.45.-a, 05r.45.Xt, 89.75.-k}
%\keywords{nonlinear dynamics,spintronics,synchronization}
\maketitle
\section{Introduction}
Since the discovery of the spin transfer torque (STT) effect by Slonczewski~\cite{slon} and Berger~\cite{berger}, tremendous and continuous interest has been shown in the study of magnetization in STNOs because of its potential applications in magnetic memory technologies~\cite{he,sharma,rana}.  Compared to the conventional STT, creating and detecting spin polarized currents in the bilayer structures, consisting of heavy metal and ferromagnetic layers~\cite{han,fukami,oh}, and trilayer structures where the ferromagnetic layers get separated by a nonmagnetic layer, are important for the study of frequency enhancement~\cite{arun_jap,arun_ieee}, energy efficiency~\cite{kubo,rehm}, fast magnetization switching~\cite{rehm} and spintronic-based neuromorphic computing systems~\cite{hoo}.  In recent times, modern spintronic devices utilize multilayer structures where the coupling between the magnetic layers has to be controlled to get enhanced magnetic properties~\cite{duine}. In an STNO the tunability in the thickness of the 3d, 4d and 5d nonmagnetic metallic spacer layer leads to ferromagnetic or antiferromagnetic coupling between the free and pinned ferromagnetic layers. This inter-layer exchange coupling has been intensively investigated since 1980s~\cite{grun,parkin,mck,omel,peng,nunn}. Recent results in this context show that the presence of exchange coupling in a magnetic layer structure is mostly applicable to modern STT magnetic random-access memory (MRAM) and sensor devices~\cite{mck,omel,peng}. 
 In particular, the study of the interlayer exchange coupling effect in synthetic magnetic multilayer systems might provide a general strategy for beyond-CMOS electronic devices~\cite{fan}.

\begin{figure}
	\centering\includegraphics[angle=0,width=0.9\linewidth]{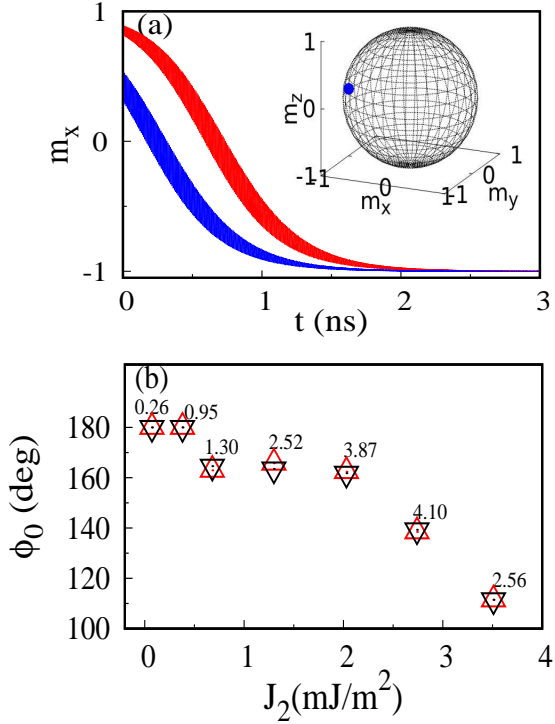}
	\caption{(a) Time evolution of $m_x$ for $I$ = 0, $J_1$ = 4.97 mJ/m$^2$ and $J_2$ = 0.919 mJ/m$^2$. (b) The experimental (Red upward triangles) and the numerical values (black downward triangles) of $\phi_0(=\cos^{-1}(m_{x0}))$ for different $J_1$ and $J_2$ corresponding to the spacer layer Ru$_{30}$Fe$_{70}$.}
	\label{fig1}
\end{figure}

Very recently, Nunn $et~al$~\cite{nunn} have reported that a noncollinear alignment between the magnetizations of the free and pinned layers can  be achieved and precisely controlled when they are coupled across a spacer layer consisting a non-magnetic material  (Ru) alloyed with a ferromagnetic material (Fe). This noncollinear alignment happens due to the biquadratic coupling which can be tuned by changing the composition and thickness of the spacer layer~\cite{nunn}. This $\mathrm{Co \left|RuFe \right| Co}$ structure is useful for the noncollinear spintronic design with magnetic moments tilted to the film plane~\cite{mck,omel,peng,nunn}. Nevertheless, there is a lack of understanding about the impact of noncollinear alignment on the dynamics of the magnetization of the free layer in an  STNO  which is yet to be understood.

Therefore, in this paper, we theoretically investigate the influence of the noncollinear alignment due to the bilinear and biquadratic couplings in a $\mathrm{Co \left|RuFe \right | Co}$~  STNO  by numerically solving the underlying Landau-Lifshitz-Gilbert-Slonczewski (LLGS) equation. Our investigations confirm the existence of switching and high-frequency oscillations of the magnetization of the free layer due to current, in the presence of the bilinear and biquadratic exchange couplings.  These studies open up  new possibilities for the applications which utilize the current induced switching and high-frequency oscillations due to interlayer exchange coupling  for the development of spintronic devices and neuromorphic computing devices~\cite{torr,mck1}. Sec. II addresses the governing equation, namely the Landau-Lifshitz-Gilbert-Slonczewski(LLGS) equation, for the magnetization dynamics of the free layer in $\mathrm{Co \left|RuFe \right | Co}$ STNO along with the bilinear and biquadratic exchange couplings. The current induced switching, oscillations of the magnetization of the free layer and its tunability by current and interlayer exchange coupling along with the expressions for the critical values of current and bilinear exchange coupling for the possibility of oscillations are presented in Sec. III. Finally, concluding remarks are made in Section IV. In Appendix A we have presented a methodology to identify the equilibrium orientation of the magnetization in the absence of the current using a microwave field.  In Appendix B, the derivation for the critical values of the current and finally in Appendix C, we have discussed the output power and its enhancement.

\section{Model}
We consider an STNO with a spacer layer Ru$_{100-x}$Fe$_x$ consisting of a nonmagnetic material (Ru) alloyed with a ferromagnetic material (Fe)~\cite{nunn}. This layer is sandwiched between two ferromagnetic layers, which includes a free layer where the direction of the magnetization can change and a pinned layer where the magnetization is fixed along the positive $x$-direction. The plane of the layers is considered to be aligned with the $xy$-plane. The dynamics of the unit magnetization vector of the free layer ${\bf m}$ is given by the LLGS equation 

\begin{align}
\frac{d{\bf m}}{dt}=&-\gamma{\bf m}\times{\bf H}_{eff}+ \alpha{\bf m}\times\frac{d{\bf m}}{dt} +\gamma H_{S}~ {\bf m}\times ({\bf m}\times{\bf p}),\label{llg} 
\end{align}

In Eq.\eqref{llg}, ${\bf m}$ = $m_x~{\bf e}_x + m_y~{\bf e}_y + m_z~{\bf e}_z$ or ${\bf m}=(m_x,m_y,m_z)$, $|{\bf m}| = 1$. Here, ${\bf e}_x,{\bf e}_y$ and ${\bf e}_z$ are the unit vectors along $x$, $y$ and $z$ directions, respectively. In spherical polar coordinates $m_x=\sin\theta\cos\phi,~m_y=\sin\theta\sin\phi$ and $m_z=\cos\theta$, where $\theta$ and $\phi$ are polar and azimuthal angles, respectively.   $\gamma$ is the gyromagnetic ratio, $\alpha$ is the damping parameter and $H_{S} = {H_{S0}}/({1+\lambda~ {\bf m \cdot p}})$, where  $H_{S0}=\hbar\eta I/2 e M_s V$ and ${\bf p}={\bf e}_x$. Here $I$ is the current passing through the free layer, $\hbar(=h/2\pi)$ is the reduced Plank's constant, $e$ is the electron charge, $M_s$ is the saturation magnetization and $V$ is the volume of the free layer. The dimensionless parameters $\eta$ and $\lambda$ determine the magnitude and angular dependence of the spin-transfer torque.
The energy density of the free layer is given by
\begin{align}
E = &\frac{J_1}{d} ~{\bf m}\cdot{\bf p} + \frac{J_2}{d} ~({\bf m}\cdot{\bf p})^2-\frac{M_s}{2}[ H_k -  4\pi M_s ]({\bf m}\cdot{\bf e}_z)^2 \label{E}
\end{align}
and the effective magnetic field ${\bf H}_{eff}=- \partial E/\partial (M_s {\bf m})$ of the free layer is
\begin{align}
{\bf H}_{eff}=(H_k-4\pi M_s) m_{z}~{\bf e}_z-\frac{1}{dM_s}(J_{1}+2J_{2} m_x)~{\bf e}_x  \label{Heff},
\end{align}
where $J_1$ and $J_2$ are the bilinear and biquadratic exchange coupling constants, respectively~\cite{mck}.  
The effective field ${\bf H}_{eff}$ includes the field due to bilinear and biquadratic couplings, magneto-crystalline anisotropy field $H_k$ and demagnetization field $4\pi M_s$.  In Eqs.\eqref{E} and \eqref{Heff}, $d$ is the thickness of the free layer.  The material parameters are considered as $M_s$ = 1210 emu/c.c., $H_k$ = 3471 Oe, $\eta$ = 0.54, $\lambda$ = $\eta^2$, $d$ = 2 nm, $A$ = $\pi\times$60$\times$60 nm$^2$, $V$ = $Ad$, $\alpha$ = 0.005 and $\gamma$ = 17.64 Mrad/(Oe s)~\cite{arun_jap,arun_ieee,nunn,tani,kubo}. Here $A$ is the cross sectional area of the STNO. In this paper, $J_1$ is restricted to be positive, corresponding to the Fe concentration below 75\% in the spacer, since oscillations are not observed for negative $J_1$.
\section{Results and Discussion}
Before applying any current the magnetization would settle in the $xy$-plane at $(m_{x0},m_{y0},m_{z0}) = (-1,0,0)$ when $J_1\geq 2J_2$ or at $\left(-J_1/2J_2,\sqrt{1-J_1^2/4J_2^2},0\right)$ when $J_1<2J_2$~\cite{nunn}. This is confirmed in Fig.\ref{fig1}(a), where the time evolution of $m_x$ is plotted for $J_1$ = 4.97 mJ/m$^2$ and $J_2$ = 0.919 mJ/m$^2$ for two different initial conditions and we can see that finally, ${\bf m}$ reaches the state (-1,0,0) (see the inset of Fig.\ref{fig1}(a)) irrespective of the initial condition. Similarly, the polar angle $\phi_0(=\cos^{-1}(m_{x0})$ since $\theta_0=\pi/2$), corresponding to $(m_{x0},m_{y0},m_{z0})$  is plotted for different sets of $J_1$ and $J_2$ in the absence of current and marked by downward black triangles in Fig.\ref{fig1}(b).   The values of $\phi_0$ can also be determined using microwave field [see Appendix A]. The red upward triangles are the experimental values measured by Nunn $et~al$  corresponding to the spacer layer Ru$_{30}$Fe$_{70}$ which confirms the validation of the macrospin simulation~\cite{nunn}.

\begin{figure}
	\centering\includegraphics[angle=0,width=1\linewidth]{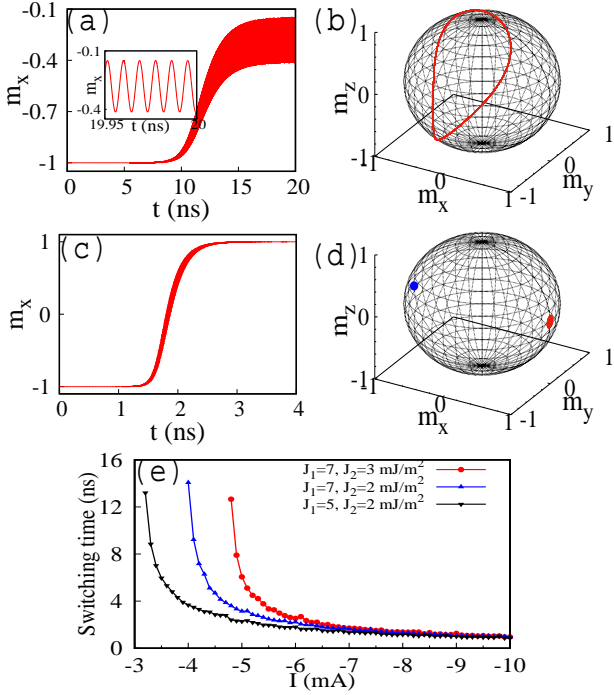}
	\caption{ Time evolution of $m_x$ and trajectory of ${\bf m}$ for (a-b) $I$ = -1.5 mA and (c-d) $I$ = -5 mA. Here, $J_1$ = 4.97 mJ/m$^2$ and $J_2$ = 0.919 mJ/m$^2$. (e) Variation of switching time versus current for different values of  $J_1$ and $J_2$. Inset in (a) corresponds to the oscillations for the time between 19.95 ns and 20 ns.}
	\label{fig2}
\end{figure}

In the presence of current, the magnetization leaves the $(m_{x0},m_{y0},m_{z0})$ state and exhibits three kinds of dynamics depending upon the values of $I$ and $J_1$ (or $J_2$). The magnetization ${\bf m}$ may continue in the same initial state, or switch from one to another state or oscillate. The magnetization ${\bf m}$ switches to $(1,0,0)$ when $|I|>|I_c^{max}|$ or oscillates when $|I_c^{min}|<|I|<|I_c^{max}|$.  When $|I|<|I_c^{min}|$ it continues to exist in $(-1,0,0)$ for $J_1\geq 2J_2$ or switches to $(m_{x1},m_{y1},m_{z1})\approx (m_{x0},m_{y0},m_{z0})$ for $J_1<2J_2$.    Here $I_c^{min}$ and $I_c^{max}$ are the critical currents between which oscillations are possible, which can be derived as [see Appendix B for more details]
\begin{align}
I_c^{max} = -\frac{\alpha eA(1+\lambda)}{\hbar \eta}\left[2 (J_1+2J_2)+(H_k-4\pi M_s) d M_s\right], \label{icmax}\\
I_c^{min} = \begin{cases} &-\frac{\alpha e A(1-\lambda)}{\hbar \eta}\left[2(J_1-2J_2)-(H_k -4\pi M_s)dM_s\right]\\&~~~~~~~~~~~~~~~~~~~~~~\rm{for}~J_1\geq 2J_2,\\
&-\frac{\alpha eA{(2J_2-J_1\lambda)}^2}{\hbar \eta J_2}\left[\frac{2J_2 d M_s(H_k-4\pi M_s)-(4J_2^2-J_1^2)}{\lambda(4J_2^2-J_1^2)-2J_2(2J_2-J_1\lambda)}\right]\\&~~~~~~~~~~~~~~~~~~~~~~\rm{for}~J_1<2 J_2.\end{cases}\label{icmin}
\end{align}
In Figs.\ref{fig2}(a-d), the time evolution and trajectories of ${\bf m}$ corresponding to the oscillations and switching are plotted for $J_1$ = 4.97 mJ/m$^2$ and $J_2$ = 0.919 mJ/m$^2$. The values of $I_c^{max}$ and $I_c^{min}$  are calculated as -2.214 mA and -1.026 mA  from Eqs.\eqref{icmax} and \eqref{icmin}, respectively. When the current is below -1.026 mA the magnetization continues to get settled at (-1,0,0) as shown in the inset of Fig.~\ref{fig1}(a). { When $I$ = -1.5 mA, the magnetization exhibits oscillations as shown in Figs.~\ref{fig2}(a-b) and when $I$ = -5 mA it switches from $m_x=$-1 to 1 as shown in Figs.~\ref{fig2}(c-d).  Note that the oscillations (inset of Fig.~\ref{fig2}(a))) are achieved due to the balance between the energy supplied by STT and the energy dissipated by damping~\cite{tani}.  The switching time, that is the time for $m_x$ to cross from -1 to +1, is plotted against current for different values of $J_1$ and $J_2$ in Fig.~\ref{fig2}(e).  The switching current, that is the current above which the switching takes place, can be obtained from Eq.\eqref{icmax}. This equation implies that the switching current increases with $J_1$ and $J_2$.  Also, we can see that  at a given value of current the switching time decreases with a decrease of $J_1$ or $J_2$. 
%The time evolution of $m_x$ while switching from -1 to +1 due to current -4 mA for $J_1$ = 5 mJ/m$^2$ and $J_2$ = 2 mJ/m$^2$ is plotted in the inset of Fig.\ref{fig2}(e).

As we discussed before, oscillations are exhibited for the currents between $I_c^{min}$ and $I_c^{max}$. To discuss the impact of the current on the frequency for different values of $J_1$ and $J_2$, current versus frequency is plotted in Fig.~\ref{fig3}(a) for $J_1<2J_2$. The oscillations of $m_x$ is shown as inset of Fig.\ref{fig3}(a) when  $I$ = -3.5 mA,  $J_1$ = 5 mJ/m$^2$ and $J_2$ = 4.5 mJ/m$^2$. From Fig.\ref{fig3}(a) we can understand that as the value of $J_1$ decreases the minimum current required to achieve the oscillations increases while when $J_2$ increases the maximum frequency attainable by the current also increases.  Also, we can observe that the frequency can be increased by the current up to or above 300 GHz, which is the largest value achieved to date for the nanopillar shaped STNOs. To investigate the influence of $J_1$ on the frequency, Fig.\ref{fig3}(b) is plotted for the frequency when $J_2$ = 5 mJ/m$^2$, where the nonoscillatory regions are represented by white color which includes two steady states $(m_{x1},m_{y1},m_{z1})$ for $J_1<2 J_2$ and $(-1,0,0)$ for $J_1\geq 2J_2$  below $I_c^{min}$, and one steady state $(1,0,0)$ above $I_c^{max}$. The critical currents $I_c^{min}$ and $I_c^{max}$, which are obtained from Eqs.\eqref{icmin} and \eqref{icmax} respectively, are plotted as open circles and triangles in Fig.~\ref{fig3}(b), respectively, match well with the numerical results. From Fig.~\ref{fig3}(b) we can observe that there is a critical value for $J_1$ where $I_c^{max}=I_c^{min}$, namely $J_1^c$, below which there occur no oscillations due to the current. The value of $J_1^c$ can be determined by equating the Eqs.\eqref{icmax} and \eqref{icmin} as
\begin{align}
J_1^c = \frac{1}{3\lambda^2}\left(2J_2 \lambda(3+2\lambda) - (Q+\sqrt{S})^{1/3} + \frac{J_2 \lambda^3 R}{(Q+\sqrt{S})^{1/3}} \right),\label{J1c}
\end{align}
where $P=(\lambda-1)dM_s(H_k-4\pi M_s)$, $Q=J_2^2 \lambda^3(-9P\lambda(3+\lambda)-4J_2(-27+\lambda(-27+\lambda(27+43\lambda))) )$, $R = 3P-16J_2\lambda$ and $S=J_2^3 \lambda^9 R^3+Q^2$.
As confirmed in Fig.~\ref{fig3}(b), the value of $J_1^c$ is  4.27 mJ/m$^2$ for $J_2$ = 5 mJ/m$^2$. For $J_1<J_1^c$, the increase in current drives the magnetization to switch from ($m_{x1},m_{y1},m_{z1}$) to $(1,0,0)$ without any oscillations.  For $J_1^c<J_1<2J_2$, the bilinear coupling decreases the value of $I_c^{min}$ and increases the value of $I_c^{max}$, implying that the tunability range of the frequency increases.
\begin{figure}
	\centering\includegraphics[angle=0,width=1\linewidth]{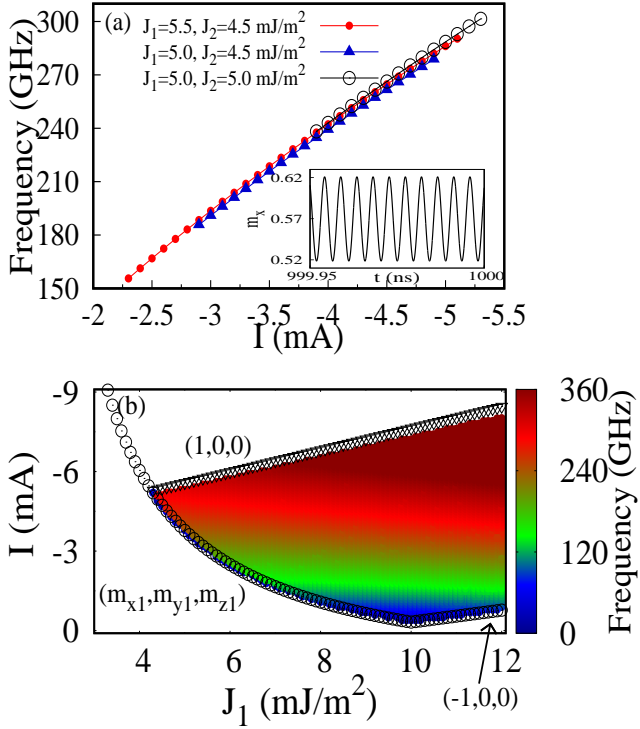}
	\caption{(a) Variation of oscillations frequency for current. (Inset) Time evolution of $m_x$ (216 GHz) for $I$ = -3.5 mA and $J_1$ = 5~mJ/m$^2$, $J_2$ = 4.5 mJ/m$^2$. (b) Frequency variation due to the current and $J_1$ for $J_2$ = 5~mJ/m$^2$.}
	\label{fig3}
\end{figure}
The output power of the STNO corresponding to $I$ = -4 mA, $d$ = 3.2 nm, $J_1$ = 5 mJ/$^2$ and $J_2$ = 4.5 mJ/$^2$ is determined as 0.41 nW. The impact of the current and the thickness of the free layer are investigated and found that they can enhance the power up to 20.91 and 1.69 times (see Appendix C).

\section{Conclusion}
In conclusion, we have investigated the dynamics of a  $\mathrm{Co \left|RuFe \right | Co}$ STNO with the bilinear and biquadratic couplings using the LLGS equation and  identified that the magnetization may continue in the same state or switch from one state to another or oscillate depending upon the magnitude of the current. We have examined the current induced switching from anti-parallel to parallel alignment through steady oscillations.  This field-free current induced switching may be helpful for low power consumption and efficient practical applications which are very robust against strong field disturbances~\cite{liu}.  We have observed that within the oscillatory region the bilinear coupling can reduce the minimum current for the oscillations and increase the tunability range of the frequency.  Similarly, the biquadratic coupling increases the maximum frequency that can be obtained by the current. We have shown that in the presence of bilinear and biquadratic couplings the frequency can be enhanced above 300 GHz by the current. We believe that the experimental achievements on noncollinear alignments in STNO may provide greater avenue for applications related to microwave generation and current induced switching.

\section*{Acknowledgments}

The works of V.K.C. and R. G are supported by the DST-SERB-CRG Grant No. CRG/2020/004353 and they wish to thank DST, New Delhi for computational facilities under the DST-FIST programme (SR/FST/PS-1/2020/135) to the Department of Physics.   M.L. wishes to
thank the Department of Science and Technology for the award of a DST-SERB National Science Chair  under Grant No. NSC/2020/00029 in which R. Arun is supported by a Research Associateship.

\section*{\NoCaseChange{Appendix A: Identification of the orientation of ${\bf m}$ in the obsence of current by using microwave field}}
Before applying any current, the magnetization would settle in the $xy$-plane at $(m_{x0},m_{y0},m_{z0}) = (-1,0,0)$ when $J_1\geq 2J_2$ or at $\left(-J_1/2J_2,\sqrt{1-J_1^2/4J_2^2},0\right)$ when $J_1<2J_2$.  The spherical polar coordinates corresponding to $(m_{x0},m_{y0},m_{z0})$ are given by $(\phi_0,\theta_0) = (\cos^{-1}(m_{x0}),\pi/2)$. In Fig.1(b) of the main text, the values of $\phi_0$ for different values of $J_1$ and $J_2$ have been plotted (black downward triangles) from the time evolution of $m_{x}$ as shown in Fig.1(a). The polar angle $\phi_0$ where the magnetization would settle before applying any current can also be determined by applying a microwave field perpendicular (along z-axis) to the plane of the free layer. Let the effective field that includes the microwave field along positive z-direction be
\begin{align}
{\bf H}_{eff}=((H_k-4\pi M_s) &m_{z}+H_a~\sin(2\pi f t))~{\bf e}_z\nonumber\\&-\frac{1}{dM_s}(J_{1}+2J_{2} m_x)~{\bf e}_x  , \tag{A1}\label{S11}
\end{align}
where the amplitude of the perpendicular microwave field $H_a$ = 100 Oe and the frequency of the microwave field $f$ = 1 GHz.  When the microwave field is applied perpendicular to the plane, the magnetization ${\bf m}$ will get perturbed about $\phi_0$ from $\phi_0 - (\triangle \phi/2)$ to $\phi_0 + (\triangle \phi/2)$, where $\triangle \phi$ is the angular range of the perturbation.  This perturbation of ${\bf m}$ can be projected along any unit vector $\hat{\bf m'} (= \cos\psi~{\bf e}_x+\sin\psi~{\bf e}_y)$ along the direction $\psi$ in the xy-plane. The component of the projected ${\bf m}$ along ${\bf m'}$ is given by $m_p = {\bf m}\cdot{\bf m'}=m_x\cos\psi + m_y\sin\psi$. The range of the projected ${\bf m}$ along ${\bf m'}$ can be determined as $\triangle m_p = m_p^{max}-m_p^{min}$, where the $m_p^{max}$ and $m_p^{min}$ are the maximum and minimum values of the projection $m_p$, respectively. Since the magnetization will get perturbed in the perpendicular direction to its initial orientation, the value of $\triangle m_p$ will be minimum when $\psi=\phi_0$ and maximum when $\psi=\phi_0$ - 90$^\circ$.  The values of $\triangle m_p$ against $\psi$ for different pair of bilinear and biquadratic exchange coupling constants $(J_1,J_2)$ mJ/m$^2$ are plotted in Fig. \ref{sfig1}.

\begin{figure}
	\centering\includegraphics[angle=0,width=1.0\linewidth]{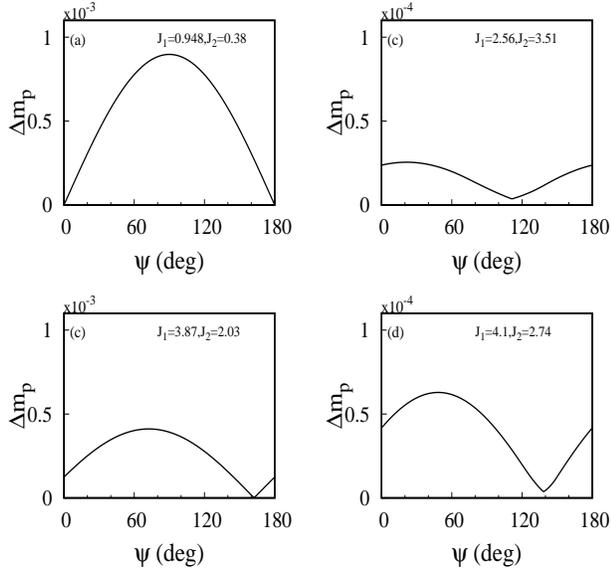}
	\caption{Range of projected ${\bf m}$ along different directions $\psi$. The plots have been plotted for different values of $(J_1,J_2)$ (mJ/m$^2$).}
	\label{sfig1}
\end{figure}

The minimum values of $\triangle m_p$ indicates the angle of orientation of the magnetization $\phi_0$ in the absence of current.  From Fig. \ref{sfig1} we can identify the values of $\phi_0$ corresponding to ($J_1$ = 2.56, $J_2$ = 3.51) mJ/m$^2$, (4.1,2.74) and (3.87,2.03) as 111$^\circ$, 138$^\circ$ and $162^\circ$, respectively. The plot corresponding to (0.948,0.38) indicates two minimum values of $\phi_0$ as 0$^\circ$ and 180$^\circ$. Since the positive value of $J_1$ leads to antiferromagnetic coupling as we mentioned in the main text, $\phi_0$ is 180$^\circ$ for $J_1$ = 0.948 mJ/m$^2$ and $J_2$ = 0.38 mJ/m$^2$. These values of $\phi_0$ exactly match with the experimental and numerical values of $\phi_0$ plotted in Fig.1(b) of the main text.

\section*{\NoCaseChange{Appendix B: Determination of critical currents}}
In this appendix, we briefly explain the procedure to determine the minimum and maximum critical currents between which the STNO exhibits the magnetization oscillations. Let us consider the LLGS equation (Eq.(1)) transformed into the spherical polar co-ordinates by using the transformation equations $m_x=\sin\theta\cos\phi,~m_y = \sin\theta\sin\phi,~m_z=\cos\theta$:
%We will briefly explain the procedure to deduce the expressions for critical currents $I_c^{max}$ and $I_c^{min}$ for $J_1<2J_2$ and $J_1\geq 2J_2$.
%The LLGS equation (Eq.(1)) can be written in spherical polar coordinates as given by
\begin{align}
\frac{d\theta}{dt} ~=~ & \frac{\gamma}{1+\alpha^2} \Bigg \{ -P(\alpha\cos\theta\cos\phi-\sin\phi)\nonumber\\ &-\alpha (H_k-4\pi M_s) \sin\theta\cos\theta\nonumber\\ &- H_{S0}\frac{(\alpha\sin\phi+\cos\theta\cos\phi)}{(1+\lambda \sin\theta\cos\phi)} \Bigg \} = M(\theta,\phi), \tag{B1}\label{S1}\\
~\frac{d\phi}{dt} ~=~ & \frac{\gamma \csc\theta }{1+\alpha^2} \Bigg \{P(\cos\theta\cos\phi+\alpha\sin\phi)\nonumber\\ &+ (H_k-4\pi M_s) \sin\theta\cos\theta\nonumber\\& + H_{S0}\frac{(\sin\phi-\alpha\cos\theta\cos\phi)}{(1+\lambda \sin\theta\cos\phi)}  \Bigg \} =N(\theta,\phi).\tag{B2}\label{S2}
\end{align}
Here $\theta$ and $\phi$ are the polar and azimuthal angles, respectively, $H_{S0} = \hbar\eta I/2eM_sV$ and $P = (J_1+2J_2\sin\theta\cos\phi)/dM_s$.
%The magnetization of the free layer settles in to the state ($\theta,\phi$)=($\pi/2,0$) i.e. (1,0,0) when the current is increased above $I_c^{max}$. Which means that the nature of the stability of the point $(\pi/2,0)$ changes from unstable to stable when the current is increased above $I_c^{max}$. 
The nature of the stability of a fixed point  ($\theta^*,\phi^*$), where $M(\theta^*,\phi^*)=N(\theta^*,\phi^*)=0$, can be identified from the eigen values of the Jacobian matrix,
\begin{align}
J = 
\begin{pmatrix}
{\left.\frac{dM}{d\theta}\right\vert}_{(\theta^*,\phi^*)} & {\left.\frac{dM}{d\phi}\right\vert}_{(\theta^*,\phi^*)}\\ {\left.\frac{dN}{d\theta}\right\vert}_{(\theta^*,\phi^*)} & {\left.\frac{dN}{d\phi}\right\vert}_{(\theta^*,\phi^*)}
\end{pmatrix}, \tag{B3}\label{S3}
\end{align}
associated with Eqs.\eqref{S1} and \eqref{S2} corresponding to ($\theta^*,\phi^*$).
%where $M(\theta,\phi)~=~\frac{d\theta}{dt}$ and $N(\theta,\phi)~=~\frac{d\phi}{dt}$. 
%The characteristic equation of $M$  can be written as
%\begin{align}
%\lambda_e^2-{\rm Tr}(M) ~\lambda_e +{\rm Det}(M) = 0,\tag{S.4}\label{S4}
%\end{align}
%where ${\rm Tr}(M)$ and ${\rm Det}(M)$ are the trace and determinant of $M$, respectively. The $\lambda_e$ is eigen value, which may be complex in general. 
The fixed point $(\theta^*,\phi^*)$ will be stable only when the real parts of both the eigen values are negative. According to Routh-Hurwitz's criterion the real parts of both the eigen values will be negative if and only if the trace of the matrix $J$ becomes negative,
%According to Routh-Hurwitz's criterion the fixed point $(\theta^*,\phi^*)$ will be stable only when the trace of $J$ satisfies the following condition.
\begin{align}
{\rm Tr}(J)<0. \tag{B4}\label{S4}
\end{align}

In the absence of the current $(\theta^*,\phi^*)=(\pi/2,\cos^{-1}(-J_1/2J_2))$ for $J_1<2J_2$ or $(\pi/2,\pi)$ for $J_1\geq 2J_2$. 

When $|I|<|I_c^{min}|$  and $J_1<2J_2$ the $(\theta^*,\phi^*)$ is approximately equal to $(\pi/2,\cos^{-1}(-J_1/2J_2))$ and becomes stable where the magnetization settles. The trace of the matrix corresponding to ($\pi/2,\cos^{-1}(-J_1/2J_2)$) is given by
\begin{align}
{{\rm Tr}(J)|}_{(\theta^*,\phi^*)} = &\frac{\gamma}{1+\alpha^2}\left[ \alpha(H_k-4\pi M_s)-\frac{\alpha(4J_2^2-J_1^2)}{2J_2 d M_s}\right.\nonumber \\&\left.+\frac{H_{S0}}{2J_2-J_1\lambda}\left(\frac{{(4J_2^2-J_1^2)}^2}{2J_2-J_1\lambda}-2H_{S0}J_1\right)\right].\tag{B5}\label{S5}
\end{align}
The minimum critical current $I_c^{min}$ (for $J_1<2J_2$), below which the fixed point $(\pi/2,\cos^{-1}(-J_1/2J_2))$ is stable, will be derived from the condition of stability \eqref{S4} and by using the Eq.\eqref{S5} as
\begin{align}
I_c^{min} = &-\frac{eA\alpha{(2J_2-J_1\lambda)}^2}{\hbar \eta J_2}\nonumber \\&\left[\frac{2J_2 d M_s(H_k-4\pi M_s)-(4J_2^2-J_1^2)}{\lambda(4J_2^2-J_1^2)-2J_2(2J_2-J_1\lambda)}\right]\tag{B6}\label{S6}
\end{align}
%\subsection*{(iii) $I_c^{min}$ for $J_1\geq2J_2$}

Similarly, when $|I|<|I_c^{min}|$  and $J_1\geq2J_2$ the state $(\theta^*,\phi^*) ~=~ (\pi/2,\pi)$ becomes stable and the magnetization settles into (-1,0,0). The trace of the matrix $J$ corresponding to the $ (\pi/2,\pi)$ is derived as
\begin{align}
{{\rm Tr}(J)|}_{(\pi/2,\pi)}  = -\frac{\gamma}{(1+\alpha^2)}&\left[\frac{2H_{S0}}{1-\lambda}-(H_k-4\pi M_s)\alpha\right.\nonumber\\&\left.+\frac{2(J_1-2J_2)\alpha}{dM_s}\right]. \tag{B7}\label{S7}
\end{align}
From the condition \eqref{S4} and Eq.\eqref{S7}, we can derive the minimum critical current (for $J_1\geq2J_2$) below which the $(\pi/2,\pi)$ is stable as
\begin{align}
I_c^{min} = -\frac{eA\alpha(1-\lambda)}{\hbar\eta}\left[2(J_1-2J_2)-dM_s(H_k-4\pi M_s)\right].\tag{B8}\label{S8}
\end{align}

When $|I|>|I_c^{max}|$ the magnetization state ($\theta^*,\phi^*$) = ($\pi/2,0$) becomes stable and the magnetization settles into (1,0,0). The trace of the matrix $J$ corresponding to ($\pi/2,0$) is 
\begin{align}
{{\rm Tr}(J)|}_{(\pi/2,0)} = \frac{\gamma}{1+\alpha^2} &\left[\frac{2(J_1+2J_2)\alpha}{dM_s}+ \frac{2 H_{S0}}{1+\lambda}\right.\nonumber\\&\left.+(H_k-4\pi M_s)\alpha \right]. \tag{B9}\label{S9}
\end{align}
The maximum critical current $I_c^{max}$, above which ($\pi/2,0$) becomes stable, can be derived from the condition \eqref{S4} and Eq.\eqref{S9} as 
\begin{align}
I_c^{max} = -\frac{\alpha eA(1+\lambda)}{\hbar \eta}\left[2 (J_1+2J_2)+(H_k-4\pi M_s) d M_s\right]. \tag{B10}\label{S10}
\end{align}

\section*{\NoCaseChange{Appendix C: Output power and its enhancement}}
The output power $P$ corresponding to the output voltage $V$ of an STNO is given by~\cite{Russ}
\begin{align}
P = \frac{V^2}{2R_L} = \frac{1}{2R_L}\left(\frac{I ~\triangle R~ R_L~({\bf m}\cdot{\bf p}) }{2(R_{avg}+R_L)}\right)^2 = \frac{I^2 ~{\triangle R}^2~ R_L~m_x^2 }{8(R_{avg}+R_L)^2}, \tag{C1}\label{S12}
\end{align}
where $\triangle R = R_{AP} - R_P$ and $R_{avg} = (R_{AP}+R_P)/2$. The quantities $R_{AP}$ and $R_P$ are the resistances while the STNO in antiparallel and parallel configuration, respectively. $R_L$ is the load resistance across which the output power $<P>_{time}$ is detected.  The time averaged power is delivered as~\cite{Russ}
\begin{align}
<P>_{time} ~=~ \frac{I^2 ~{\triangle R}^2~ R_L }{8(R_{avg}+R_L)^2} ~<m_x^2>_{time}. \tag{C2}\label{S13}
\end{align}
The expression for $R_{AP}$ can be calculated from the GMR ratio $\triangle R/R_{P}$ as $R_{AP} = R_P + (\triangle R/R_P) R_P$. Since the magnetoresistance for Co$\vert$RuFe$\vert$Co is not available in the literature, we consider the GMR ratio 0.004 corresponding to Co(3.2nm)$\vert$Ru(0.5nm)$\vert$Co(3.2nm) to find the approximated value of $<P>_{time}$ for our system~\cite{Rah}. For $R_P$ = 100 $\Omega$, $R_L$ = 50 $\Omega$ and $I$ = -4 mA, $<P>_{time}$ is determined from $<m_x^2>_{time}$ corresponding to the $m_x(t)$ between 950 ns and 1000 ns as 0.41 nW. The $<P>_{time}$ is plotted for the different currents in Fig. \ref{sfig2}(a) for $d$ = 2 nm  and different thicknesses of the free layer in Fig. \ref{sfig2}(b) for $I$ = -4 mA  when $J_1$ = 5 mJ/m$^2$ and $J_2$ = 4.5 mJ/m$^2$. From Fig. \ref{sfig2}(a) and (b) we can understand that the power can be enhanced by the current up to 20.91 times and the free layer's thickness up to 1.69 times corresponding to  $J_1$ = 5 mJ/m$^2$ and $J_2$ = 4.5 mJ/m$^2$, which confirms the tunability and enhancement of power by the current and the thickness of the free layer.  In general, the output power can be enhanced further by connecting the STNOs parallelly or serially and phase locking them by microwave field and current~\cite{Kaka,Leb}.
\begin{figure}
	\centering\includegraphics[angle=0,width=1.0\linewidth]{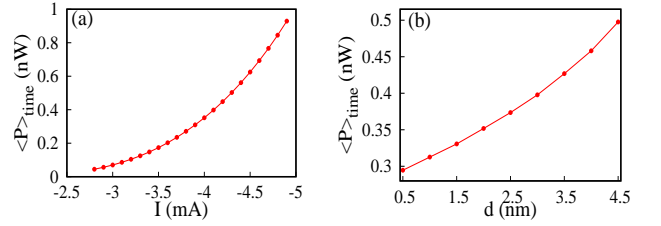}
	\caption{The variation of time averaged power against (a) current for $d$ = 2 nm and (b) thickness of the free layer for $I$ = -4 mA. Here, $J_1$ = 5 mJ/m$^2$ and $J_2$ = 4.5 mJ/m$^2$.}
	\label{sfig2}
\end{figure}

\section*{AUTHOR DECLARATIONS}
\subsection*{Conflict of Interest}
The authors have no conflicts to disclose.
\subsection*{DATA AVAILABILITY}
The data that support the findings of this study are available from the corresponding author upon reasonable request.

\end{document}